\documentclass[12pt]{article}

\usepackage{epsfig}

\begin{document}
\thispagestyle{empty}

\begin{center}
{\Large\bf The Riemann Zeta Function and Vacuum Spectrum\footnote{Talk at the
{\it ``Fourth International Winter Conference on Mathematical Methods in Physics'', held at 
Centro Brasileiro de Pesquisas Fisicas, Rio de Janeiro, Brazil,}   09 - 13 August 2004.  
Published in Proceedings of Science, PoS (WC2004) 026, see http://pos.sissa.it/}}

\vspace{0.5cm}

Sergio Joffily

 CBPF, Rio de Janeiro, Brazil\\ E-mail: {joffily@cbpf.br}

\end{center}

\abstract{A variant for the Hilbert and Polya spectral interpretation of 
the Riemann zeta function is proposed. Instead of looking for a self-adjoint 
linear operator H, whose spectrum coincides with the Riemann zeta zeros, 
we look for the complex poles of the S matrix that are mapped into the 
critical line in coincidence with the nontrivial Riemann zeroes. 
The associated quantum system, an infinity of ``virtual resonances" described 
by the corresponding S matrix poles, can be interpreted as the quantum vacuum. 
The distribution of energy levels differences associated to these resonances
shows the same characteristic features of random matrix theory.}

\newpage
\setcounter{page}{1}
\section{Introduction}
\paragraph*{ }

There is a conjecture, quoted to be made by 
Hilbert and Polya, that the zeros of the Riemann zeta function $
\zeta (z)$ on the critical line are eigenvalues of a ``mysterious" complex Hermitian
operator H. We report in this presentation a variant of the 
above conjecture where we 
associate the unknown quantum system 
to the ``vacuum". This vacuum is interpreted as an infinity of ``virtual resonances", described by complex poles of the scattering S matrix.

Actually the Riemann zeta zeros fluctuations are interesting by their universality, being observed in quantal spectra of different physical systems (see Refs.\cite{Mehta}), 
and by the connection they have with chaotic dynamics, (for a review see Bohigas \cite{Bohigas}). Ever since Montgomery's \cite{montgomery} conjectured these zeros behave like the eigenvalues of a random hermitian matrix,
attempts have been made in order to find a quantum system with
the Hamiltonian represented by such an operator (see Berry and Keating \cite
{berry}, and references therein).
Our goal here is to present a new conjecture in order 
to relate the origin of this universality to the statistical fluctuation 
of the vacuum that provides the same bakground for the different physics systems in 
interaction with it. 

We will begin by a short outline of the concepts of the Riemann 
zeta function, prime numbers and the Jost function. After, motivated by recent
work \cite{joffily} in which was shown numerically a one to one connection between 
the large zeros of Jost function in the complex momenta plane with large prime numbers
and large complex Riemann zeta zeros, we propose a 
distribution for these Jost zeros representing the quantum vacuum. Finally, rather
than look for a potential that put these Jost zeros on the desired place, we show
that the corresponding resonances energies has the same statistical fluctuation 
given by the random matrix theory (RMT). We conclude suggesting an approximate 
model for the above quantum vacuum.

\section{Prime numbers, Riemann's zeta  function and Jost function}
\paragraph*{ }

A comparison between prime numbers, Riemann's zeta  function and Jost function
has been given in a recent paper \cite{joffily}. In this section we summarize it 
in order to introduce the proposed map (\ref{map}) between complex zeros of 
Jost function with the complex zeros of the zeta function. 
For a recent review concerning the relevance of prime numbers to 
Physics see Rosu \cite{rosu}.

The approximate number of primes $\pi (x)$ less than a given 
$x$, also called the prime counting function, is given by the prime number
theorem $\pi (x)\sim x/\log x $ (see Titchmarsh \cite{doze}, Chapter- III). 
This relation gives the asymptotic
approximation for $n$th prime $p_{n}$, 
\begin{equation}\label{primo}
p_{n}=n\log \ n\quad ,\qquad \mbox{as}\ n\rightarrow \infty \ .
\end{equation}

The connection between the distribution of prime numbers $\pi (x)$ and the 
complex zeros of the zeta function started with Riemann's 1859 paper (see Edwards
 \cite{treze}, p. 299). Riemann's zeta function is defined (Ref.\cite{doze}, p.1),
either by the Dirichlet series or by the Euler product 
\begin{equation}\label{zeta}
\zeta (z)=\sum_{n}n^{-z}=\prod_{p}(1-p^{-z})^{-1}\ ,\qquad Re\ z >1\ ,
\end{equation}
where n runs through all integers and p runs over all primes. This function can be 
analytically continued to the whole complex plane, 
except at $z=1$ where it has a simple pole with residue 1. It
satisfies the functional equation 
$\zeta (z)=2^{z}\pi ^{z-1}\sin (\pi z/2)\Gamma (1-z)\zeta (1-z)\ $,
called the reflection formula, where $\Gamma (z)$ is the Euler gamma function. It
is known that $\zeta (z)$ has simple zeros at points $z=-2n,\
n=1,2,\dots $, which are called trivial zeros, and an infinity of
complex zeros lying in the strip $0 < Re\ z < 1$. Due to the reflection
formula they are symmetrically situated with respect to the axis $Re\ z=1/2$, 
and since $\zeta (z^{*})=\zeta ^{*}(z)$, they are also symmetric about
the real axis. Consequently it suffices to consider the zeros in the upper half of 
strip $1/2\leq Re\ z < 1$. It is possible to enumerate these
complex zeros as $ z_{n}=b_{n}+i\ t_{n}$ , with $ t_{1}\leq
t_{2}\leq t_{3}\leq \dots$, and the following result can be proven
(see Titchmarsh \cite{doze} , p.214) 
\begin{equation}\label{zass}
|z_{n}|\sim t_{n}\sim \frac{2\pi n}{\log n}\\ ,\qquad \mbox{as}\ n\rightarrow \infty .
\end{equation}
The Riemann Hypothesis (RH) is the conjecture, not yet proven, that all 
complex zeros of $\zeta (z)$ lie on the axis $Re\ z=1/2$, called the
``critical line''. This is usually considered as the most important unsolved
problem in Mathematics. Based on this conjecture, Riemann derived an exact formula 
for $\pi (x)$ taking into account 
local prime fluctuations in terms of nontrivial zeta zeros 
(Ref.\cite{treze}, p.299). 

On the other hand, the Jost function 
has played a central role in the
development of the analytic properties of the scattering amplitudes. 
In order to recall its
properties, let us consider the scattering of a
non-relativistic particle, without spin, of mass $m$ by a spherically
symmetric local potential, $V(r)$, everywhere finite, behaving at infinity as 
$V(r)=O(r^{-1-\epsilon }),\quad \epsilon >0.$
The Jost functions $f_{\pm }(k)$ are defined (see Newton \cite{quinze}, p.341) as the
Wronskian W, $f_{\pm }(k)=W[f_{\pm }(k,r),\ \varphi (k,r)]$, 
where $\varphi (k,r)$ is the regular solution of the radial Schr\"{o}dinger
equation 
\begin{equation}\label{schr}
{\lbrack }\frac{d^{2}}{dr^{2}}+k^{2}-V(r){\rbrack }\ \varphi (k,r)=0\ ,
\end{equation}
(in units for which $\hbar=2m=1$) $k$ being 
the wave number and the Jost solutions, $f_{\pm }(k,r)$, are two linearly independent
solutions of eq.(\ref{schr}). They satisfy the boundary conditions, 
$\lim_{r{\rightarrow \infty }}[e^{{\mp }ikr}f_{\pm }(k,r)]=1$, 
corresponding to incoming and outgoing waves of unit amplitude.

The properties of the solution of the differential equation (\ref{schr}) 
define the domain of analyticity of the Jost
functions $f_{\pm }(k)$ on the complex $k$-plane as well as its symmetry
properties, such as $f_{+}^{*}(k^{*})=f_{-}(k)$.
The phase of the Jost function is just
minus the scattering phase shift $\delta (k)$, that is 
$f_{\pm }(k)=\vert f_{\pm }(k)\vert \ \ e^{ {\mp }i\delta (k)}\ ,$ 
so that the usual S matrix is given by 
\begin{equation}\label{matrixs}
S(k)\equiv e^{2i\delta (k)}=\frac{f_{-}(k)}{f_{+}(k)}\ .
\end{equation}
The complex poles ($Re\ k \neq 0$) of $S(k)$, or zeros of $f_{+}(k)$,
correspond to the solutions of the Schr\"{o}dinger equation with purely
outgoing, or incoming, wave boundary conditions. Resonances show up as
complex poles with negative imaginary parts, their complex energies being 
\begin{equation}\label{energia}
k_{n}^{2}=\Xi _{n}-i\ \frac{\Gamma _{n}}{2} ,
\end{equation}
where $\Xi _{n}$ and $\Gamma _{n}$ represents the energy and the width, 
respectively, associated with $n$th resonance state in order of distance from origin. 
For small $\Gamma _{n}$, resonances appear as long-lived
quasistationary states populated in the scattering process. If the width is
sufficiently broad no resonance effect will be observed and we will
call this kind of S matrix pole as ``virtual resonances'' throughout.
Resonances, in fact, are represented by pairs 
of symmetrical S matrix poles in the complex k-plane, a capture state pole in 
the third quadrant and the decaying state in the fourth quadrant, since they 
give to the asymptotic solution an incoming growing wave and an outgoing decaying 
wave, respectively, exponential in time \cite{dezesete}. This
could be described by Gamow vectors treated by Bohm and Gadella \cite{vsete}
as pairs of S matrix poles corresponding to decay and growth states, where the traditional
Hilbert space description is replaced by the generalized Rigged Hilbert Space
in order to account for time asymmetry of a resonant process. The transient 
states which last only a very short time corresponding to the above ``virtual resonances", 
to which our instruments are insensitive, will be called here as ``vacuum". Examples of
broad resonances are the well known large poles, related basically to the
cutoff in the potential without any physical interpretation (see
Nussensveig \cite{dezoito}, p.178).

In truncating the potential, i.e., the potential is set equal to zero for $r\geq
R>0 $, which is the cutoff of the potential at arbitrarily large distances $R$, it is 
possible to obtain explicit asymptotic formulas for the Jost zeros. With
this restriction it can be shown that equation $f_{+}(k)=0$ 
is entire of order $1/2$ and according to Piccard's theorem has infinitely many roots for
arbitrary values of the potential. 
The asymptotic expansion of $k_{n}$, for large n, after introducing 
dimensionless parameter $\beta =kR$, is given by (Ref.\cite{quinze}, p.362) 
\begin{equation}\label{humblet}
\beta _{n}=n\pi \\ -i\frac{(\sigma +2)\log\vert n\vert}{2}+0(1)
\end{equation}
where $n=\pm 1,\ \pm 2,\ \pm 3\cdots $ and $\sigma $ is defined by the
first term of the potential asymptotic expansion, near $r=R$, through $%
V(r)=C(R-r)^{\sigma }+\cdots $, $\sigma \geq 0$ and $r\leq R$.

The connection between the complex zeros of the Jost function and those one
of the Riemann zeta function is provided by the transformation, 
\begin{equation}\label{map}
z=-i\frac{\beta ^{2}}{2\ Im\ \beta ^{2}}\ ,
\end{equation}
by which the lower half of complex $\beta $-plane ($Re\ \beta \neq 0$) is
mapped onto the critical axis, $Re\ z=1/2$, of complex $z$-plane. 

Now we show that for cutoff potentials the transformation (\ref{map}) gives rise to 
complex Jost zeros with the same asymptotic behavior as the complex Riemann zeta
zeros, being all in the critical line. After introducing dimensionless
quantities, energy $E_{n}=R^{2}\Xi _{n}$ and widths $G_{n}=R^{2}\Gamma _{n} $
, the equation (\ref{energia}) is written as $\beta_{n} ^{2}=E_{n}-iG_{n}/2$, then by
(\ref{map}) we get 
\begin{equation}\label{linha}
z_{n}=\frac{1}{2}+i\frac{\ E_{n}}{G_{n}}\ ,
\end{equation}
from (\ref{humblet}), with $\sigma =0$, we see that $\{Im\ z_{n}\}$ has the same asymptotic
expansion as $\{t_{n}\},$ given by (\ref{zass}), i.e., 
\begin{equation}\label{linhass}
\frac{\ E_{n}}{G_{n}}\ =\frac{t_{n}}{8}\quad \mbox{as}\quad n\rightarrow \infty \ ,
\end{equation}
which means that for each resonance, the ratio between the energy and the width is 
given by the height of the zeta zero on the critical line. On the other hand, 
dimentionless widths $\{G_{n}\}$, defined in (\ref{linha}) as 
\begin{equation}\label{width}
G_{n}=4\ Re\ \beta _{n}\ \ Im\ \beta _{n}\ ,
\end{equation}
after taking into account (\ref{humblet}), when $\sigma =0$, shows the same
asymptotic expansion for large primes (\ref{primo}), given by the prime number
theorem, 
\begin{equation}\label{widthass}
G_{n}=4\pi p_{n}\quad \mbox{as}\quad n\rightarrow \infty .
\end{equation}
Then $n$th large complex Jost zeros are also related to $n$th large
primes, showing an asymptotic connection between primes and complex Riemann
zeta zeros, in a one-to-one correspondence.

\section{Hilbert-Polya conjecture }
\paragraph*{ }

The Hilbert-Polya conjecture is the spectral interpretation of the complex zeros 
of the Riemann zeta function as eigenvalues of a self-adjoint linear operator H in 
some Hilbert space. Such H could prove the RH.
We suggest a variant for the above conjecture:
Instead of looking for H, whose spectrum have to coincide with the Riemann 
zeta zeros, we are looking for
a potential that gives a Jost function with all zeros on the lower half of 
complex $\beta $-plane ($Re\ \beta \neq 0$) that coincide with the
complex zeros of the Riemannn zeta function after the transformation (\ref{map}). In this
way the Riemann hypothesis follows. For real potentials, these complex $\beta $ 
zeros are located symmetrically about the imaginary axis, then by (\ref{map}) they
will be mapped symmetrically about the real axis into the critical line. 
The associated quantum system could be identified with the quantum vacuum, interpreted as an 
infinity of  ``virtual resonances", described by corresponding S matrix poles.

The first to associate the Riemann Hypothesis 
with transient states were Pavlov and Faddeev \cite{Faddeev}, by relating 
the nontrivial zeros of the zeta function to the
complex poles of the scattering matrix of a particle on a surface of
negative curvature. Khuri \cite{Khuri} has recently 
proposed a modification to the inverse scattering problem in order to obtain the 
potential whose coupling constant spectrum coincides with the Riemann zeta zeros.

Assuming that the RH is true, we conjecture a vacuum spectrum with widths given by
$G_{n}=p_{n}$ , according to (\ref{widthass}), 
and the corresponding energies given by
$E_{n}= G_{n}t_{n}= p_{n}t_{n}$, according to (\ref{linhass}).
Before trying to find a potential that will give these S matrix poles we will study numerically 
the statistics fluctuation of these ``virtual levels". 

\begin{figure}
\epsfxsize=7.5cm
\epsfbox{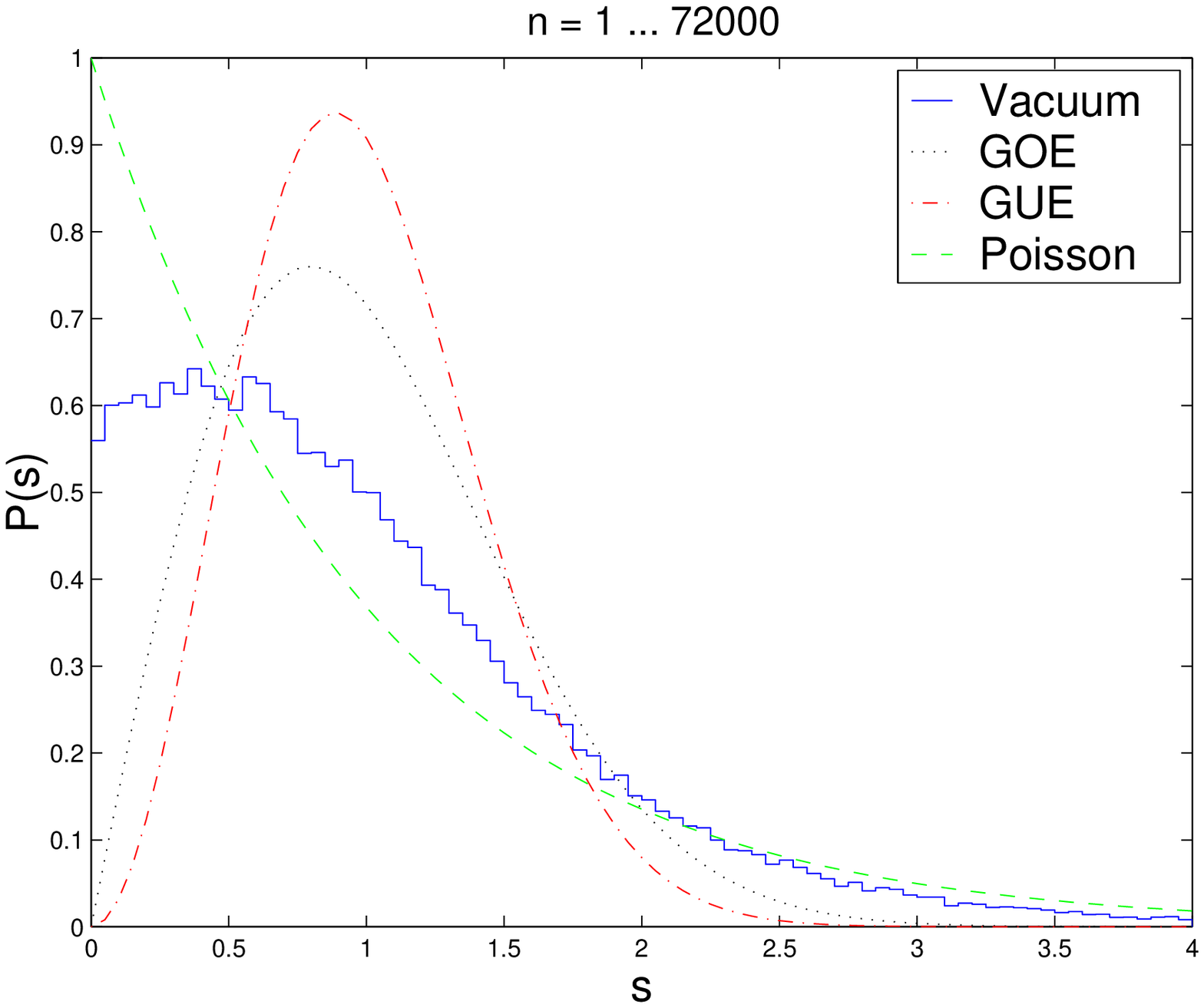}
\epsfxsize=7.5cm
\epsfbox{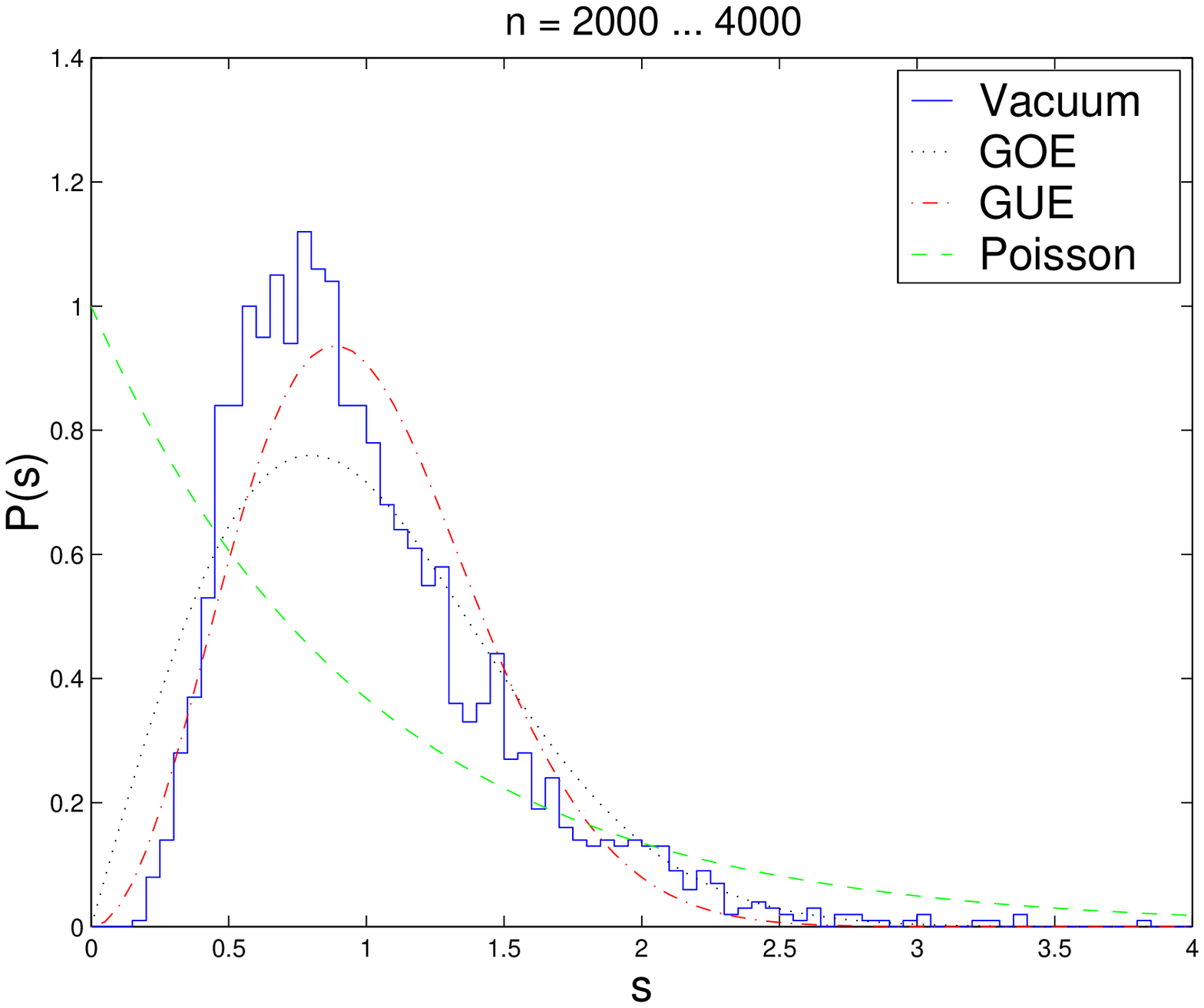}
\epsfxsize=7.5cm
\epsfbox{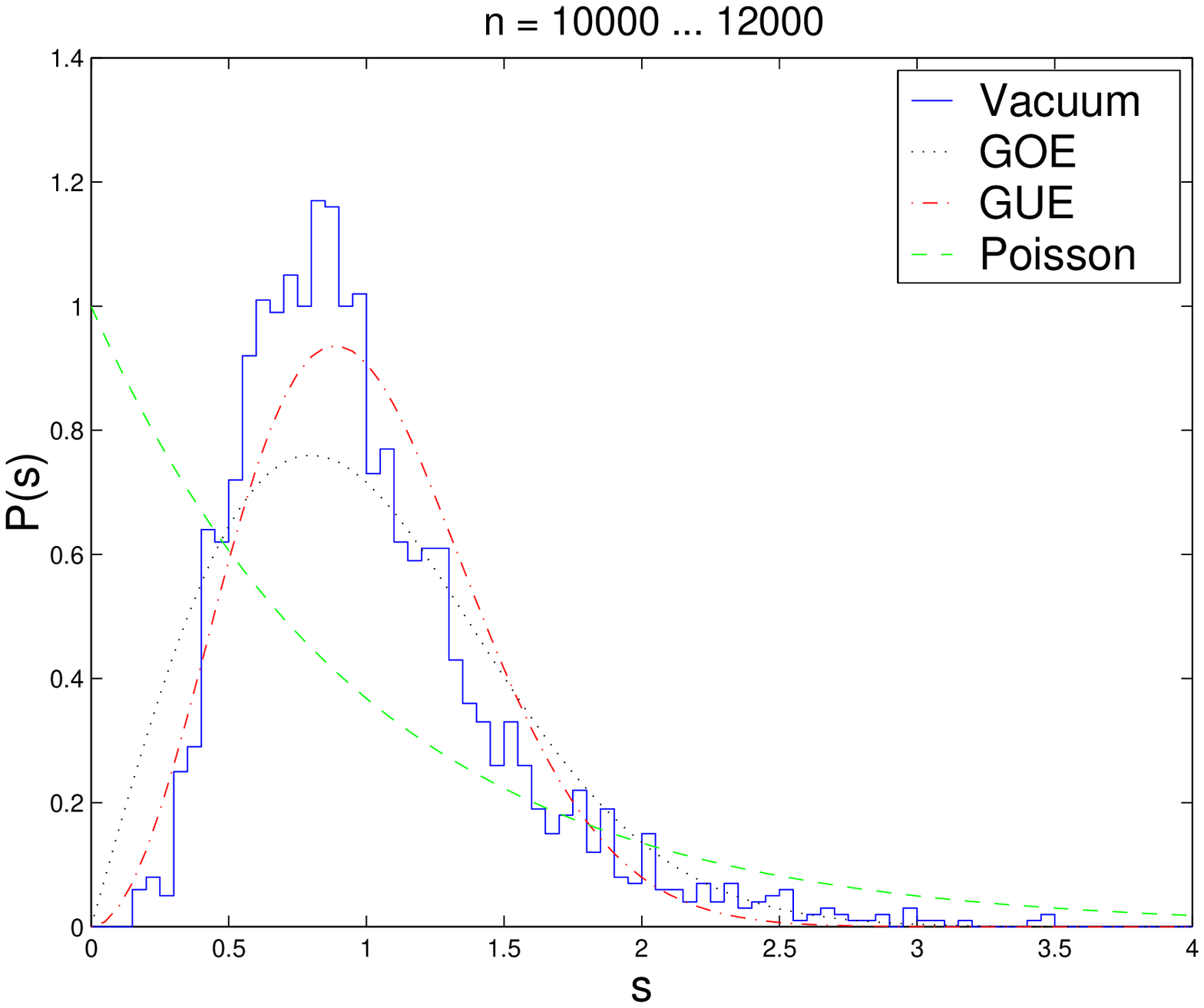}
\epsfxsize=7.5cm
\ \  \ \ \ \ \ \ \ \  \epsfbox{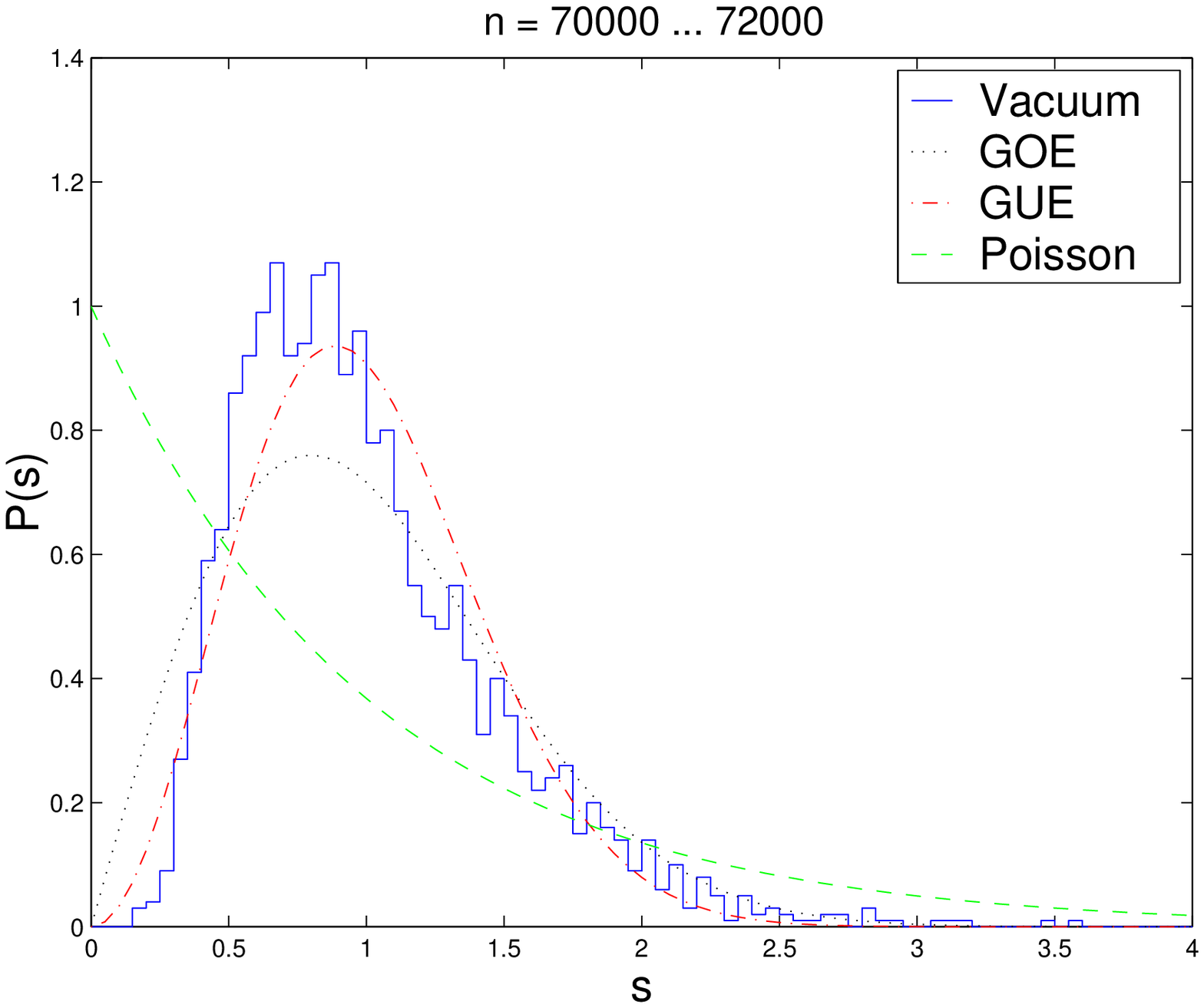}
\caption{ The statistical of level-spacings in vacuum spectra $\{p_{n}t_{n}\}$. 
The histogram 
at the top left gives a sequence comprising the first 72000 vacuum levels.  
The histogram at the top right comprises 2000 consecutive levels in the analysis corresponding
 from the 
2000$th$ to the 4000$th$ level. The histograms at the bottom give 2000 consecutive levels centered
at different $E$ values.}\end{figure}

\section{RMT and energy distribution of virtual resonances }
\paragraph*{ }

The random matrix theory (RMT) describes fluctuation properties of quantal spectra.
In lack of analytical or numerical methods to obtain the spectra of complicated Hamiltonians, 
Wigner and Dyson (see Refs. \cite{Mehta}) 
analyzed ensembles of random matrices, in which the matrix elements are considered to be 
independent random variables, that have in common only 
symmetry properties like hermiticity and time-reversal invariance.
The RMT predicts, for the nearest-neighbour spacings of the energy levels distribution $P(s)$,  
the form which is described by the Wigner surmise: $P (s) =  A~s^{D}~  exp(- Bs^{2} ) $, 
where  $s_{n} = (E_{n+1} - E_{n})$ are spaces between adjacent levels of the
unfolded spectrum with mean distance $<s_{n}> = 1 $. 
It is obtained by accumulating the number of spacings
that lying between ($s, s+ \triangle s$) and then normalizing $P (s)$ to unity. 
$A$ and $B$ are normalizing constants and $D$ is a parameter which depends on the symmetry of 
the system which characterizes the repulsion between neighbor levels. 
For hermitian time-reversal Hamiltonians 
(Gaussian Orthogonal Ensemble) $D=1$ and $P_{GOE} (s) = (\pi /2)~ s~  exp (- \pi  s^{2} / 4 )  $. 
For complex hermitian matrices, when time-reversal invariance does not hold (Gaussian Unitary Ensemble) 
$D=2$ and  $P_{GUE} (s) = ( 32/\pi^2)~ s^{2} ~ exp (- 4 s^{2} / \pi)$.                                  
If the eigenvalues of a system are completely uncorrelated (Poisson Ensemble ) 
we have $P_{PE} (s) =   exp (- s )$. 
                             
The nearest-neighbor spacing distribution $P(s)$ corresponding to the 
vacuum spectrum energy, i.e. resonances energies given by $\{p_{n}t_{n}\}$, where
prime sequence $\{p_{n}\}$ are taken from the table \cite{vcinco} and $\{t_{n}\}$ computed by 
Odlyzko \cite{vquatro} are shown in Figure 1. The curves corresponding to Poisson 
distribution (dashed line), GOE distribution (dotted line) and GUE distribution (dotted-dashed line)
are drawn for comparison. For a 
sequence comprising the first 72 000 vacuum levels we obtain, see the top left  
of Figure 1, a mixing between Poisson and GOE distribution. In order to compare 
with spectra of atomic nuclei at higher energies, in regions of high density, we need an interval 
of energy $\triangle E$ centered at $E$ like $E >> \triangle E >> 1$.     
As can be seen from the numerical results, see the top right and the bottom of Figure 1, 
when $E >> \triangle E $, the nearest-neighbor spacing distribution of the vacuum spectrum 
are consistent with a kind of mixing of GOE distribution and GUE, with the same 
characteristic features of random matrix, i.e. level repulsion 
at short distances and suppression of fluctuations at large distances. 
The histograms at the bottom of the Figure 1 gives 2000 consecutive levels centered at
differents $E$ values in order to show the spectral rigidity, i.e. a very small fluctuation
around its average. 

\section{A model for the quantum vacuum}
\paragraph*{ }

Following \cite{joffily} we present a model that approximates the above quantum vacuum. 
It could be represented by the 
non-relativistic s-wave scattering by a spherically symmetric 
potential barrier, $V(r)= V_{0}$ for $ r < R $ and $ V(r)= 0 $ for $r \geq R $.  
From the stationary scattering solution one obtains the
Jost function $f_{+}(k)$ and after 
introducing dimensionless parameters, $ \beta =kR$ and 
$v=V_{0}R^{2}$, its zeros are given by the solutions of
the complex transcendental equation 
$\sqrt{\beta ^{2}-v}\ \cot \sqrt{\beta ^{2}-v}=i\beta$, 
for each value of potential strength $v$. The displacements of the roots, 
$\beta _{n}$, in
the $\beta $-plane with the variation of the potential strength was shown many years ago by
Nussensveig \cite{vdois}. All these zeros are in the lower half of complex $%
\beta $-plane and located symmetrically about the imaginary axis. 
The energy/width ratios $\{4\pi E_{n}/G_{n}\}$ where calculated (see \cite{joffily}, Fig.1a ) 
from the Jost zeros $\beta _{n}$, for $v=2$, 
in order of distance from the origin and show an 
approximate agreement from the beginning with $\{t_{n}\}$, for n up
to $6 \times 10^{4}$.
The same Jost zeros $\beta _{n}$, for $v=2$, after
transformation (\ref{width}), gives the dimensionless widths $\{G_{n}/4\pi \}$  
in agreement with the global behavior
of the sequence of primes, from the beginning in the same range (see \cite{joffily}, Fig. 1b ). 
The local behavior, defined as deviations from the average density of the
zeta zeros, are not obtained by this potential \cite{joffily}.

The qualitative agreement of the above model
suggests the investigation of the many-body problem described by one particle being scattered by
an effective cutoff potential where, by analogy to the mean field, 
a kind of residual interaction is introduced phenomenologically, in
order to describe the ``vacuum" fluctuation with the same statistical
distribution of the vacuum energy levels $\{p_{n}t_{n}\}$. 
In this case, the universality of level fluctuation law of the 
spectra of different quantum systems (nuclei, atoms and molecules)
 \cite{Mehta,Bohigas} could be understood by the ``vacuum" role as a dissipative system 
\cite{Callen}.

\section{Conclusion}
\paragraph*{ }

In summary, the spectral interpretation 
of the Riemann zeta function is associated to the vacuum spectrum 
by means of a variant of the Hilbert-Polya conjecture. The distribution of 
the virtual resonances would reflect a chaotic nature of the quantum vacuum. 
The energy/width ratios of the virtual resonances are associated with the nontrivial 
zeta zeros while the corresponding 
widths are related to the prime sequence. Finally, 
a weak repulsive cutoff potential is proposed as an approximate model for the quantum vacuum. 

\section*{Acknowledgements}
\paragraph*{ }

It is a pleasure to thank C. A. Garcia Canal for useful conversations.

\end{document}